# Swarm UAVs Communication

Arindam Majee, Rahul Saha, Snehasish Roy, Srilekha Mandal, Sayan Chatterjee

Jan 6, 2023

# <u>INDEX</u>

# List of Figures

# <u>List of Tables</u>

# 1. Problem Statement: With the advent of Cyber-Physical Systems, IoT and 5G the need for an efficient antenna system is indispensable. Other than the circuit level designs the main key factor for the realisation of the 'smart' technologies mainly relies on the development and design of highly efficient signal transmitting and receiving units. In the civil domain, during a big calamity, there is a need for an immediate picturization of the site(s) to support the rescue forces in making decisions. The search for buried people after building collapses, search for people grouped under a temporary shelter during disaster or fire at big factories or chemical plants are possible scenarios. UAVs form a more efficient solution to mitigate this problem. In recent times, unmanned aerial vehicles (UAVs) complement ground-based searchers by being able to cover more area at far greater speed. However, micro UAVs exhibit limitations due to their size. Their payload is usually only a few hundred grams allowing only light and compact sensors to be deployed.

Accordingly, calculation of total load including electronic steering components are challenging issues and there has been no option of a secondary system to be mounted during the malfunctioning of elements.

Problems identified:

1. Malfunctioning of overall cooperative communication among drones if a single drone produces wrong radiation pattern
2. The options are to cancel the situation and call back all drones or two that have a secondary system mounted during faulty conditions. Thus weight may increase.

Outcomes in terms of application;

a. Faster location identification in a disaster management situation

b. Tracking of population movement in coastal areas pre-distributed during the cyclonic storm

c. Accurate and lightweight forest intruder detection

d. Air traffic rescue operation in hilly areas where micro UAV in the swarm can move at a closer height.

## 2. Abstract:

The advancement in cyber-physical systems has opened a new way in disaster management and rescue operations. The usage of UAVs is very promising in this context. UAVs, mainly quadcopters, are small in size and their payload capacity is limited. A single UAV can't traverse the whole area. Hence multiple UAVs or swarms of UAVs come into the picture managing the entire payload in a modular and equiproportional manner. In this work we have explored a vast topic related to UAVs. Among the UAVs quadcopter is the main focus. We explored the types of quadcopters, their flying strategy,their communication protocols, architecture and controlling techniques, followed by the swarm behaviour in nature and UAVs. Swarm behaviour and a few swarm optimization algorithms has been explored here. Swarm architecture and communication in between swarm UAV networks also got a special attention in our work. In disaster management the UAV swarm network must have to search a large area. And for this proper path planning algorithm is required. We have discussed the existing path planning algorithm, their advantages and disadvantages in great detail. Formation maintenance of the swarm network is an important issue which has been explored through leader-follower technique. The wireless path loss model has been modelled using friis and ground ray reflection model. Using this path loss models we have managed to create the link budget and simulate the variation of communication link performance with the variation of distance.

# 3. Introduction:

An unmanned aerial vehicle (UAV) is an aircraft with no onboard pilots. The emergent civilian, military, commercial, agricultural and scientific applications of Unmanned Aerial Vehicles are abundant in this world. Recently UAVs are highly used in Search and Rescue (SAR), 5G communication. UAVs can be used in various civil applications due to their ease of deployment, low maintenance cost, high mobility, and ability to hover [1]. In civil applications UAVs can be used in surveillance, agriculture (crop monitoring, pesticide spraying, etc) [2], transportation [3], traffic monitoring [4], etc. In [5] the authors have already surveyed the vast application fields of UAVs.

However, a single UAV faces a lot of issues like stability, survivability, reliability, etc [6]. A group of UAVs can perform better than a single UAV. The group of UAVs is commonly known as Swarm UAV. We will discuss swarm UAV in the upcoming sections.

We have classified the whole report into several sections. The first section describes various types of UAVs (our focus is mainly on Quadrotors), how a UAV flies, the effect of wind on the UAV and the mathematical model of its control.

## SECTION-1

# 4. Introduction to UAVs:

In this section, we are going to provide a brief idea about the UAVs, types of UAVs, how they fly and related topics.

    a) <u>Types of UAVs:</u> There are various types of aerial drones. But the most common four types are, Multi-rotor, Fixed-wing, Single rotor, and Fixed wing hybrid. The advantages, disadvantages, and main features are listed below [7-9].

| Features | Multi-rotor | Fixed Wing | Single rotor | Fixed-Wing Hybrid |
|----------|-------------|------------|--------------|-------------------|
| Flight Time | Short. Normally half an hour. | Very long. More than 8 hours. | Moderate | Long |
| Speed | slow | fast | slow | medium |
| Cost | low | high | high | high |
| VTOL* And hovering | possible | impossible | possible | possible |
| Energy efficiency | Not efficient | efficient | efficient | efficient |
| Importance of runaway | not required | required | not required | not required |
| Payload capacity | light | light | heavy | heavy |

Table 1: Types of UAVs

*VTOL stands for Vertical Take-Off and Landing.

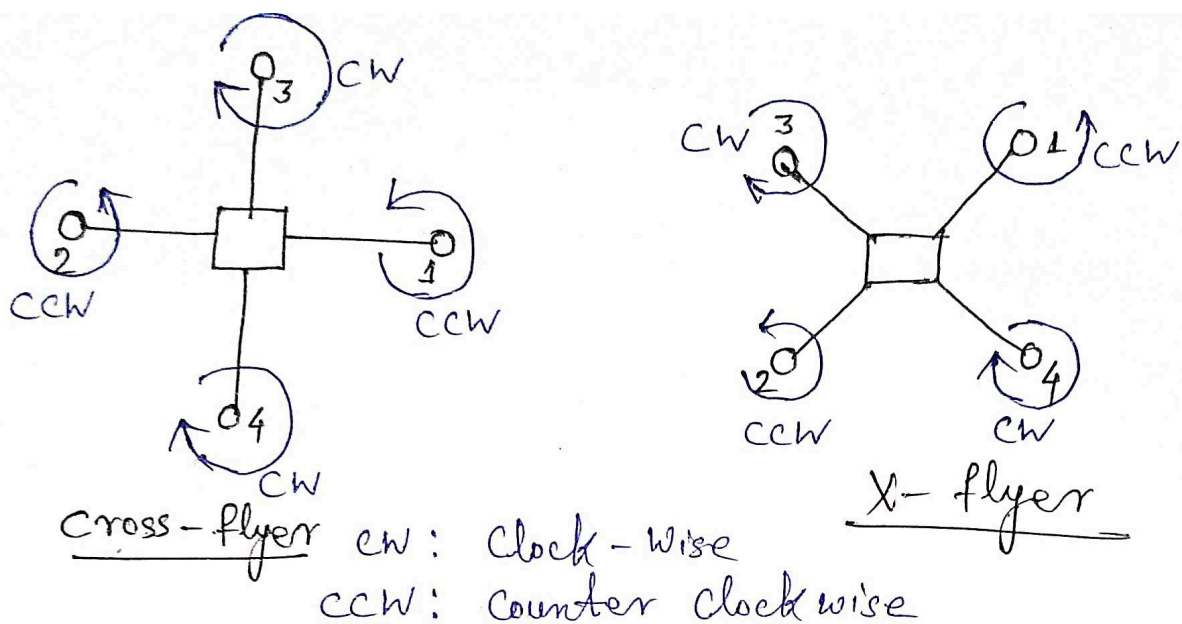

Figure 1: Quadcopter configuration

In our work, we will use quadrotors as they are flexible for VTOL, effective in terms of cost, battery life and control. A quadcopter has 4 wings that are connected to four rotors. Two rotors (each is in the other's corner) rotates in a clockwise direction and another two in the anti-clockwise direction. A quadcopter has 6 degrees of freedom among them 3 are transitional and 3 are rotational.

The quadcopter can move in 3 directions commonly known as x, y and z in a cartesian coordinate system. And the quadcopter rotates in 3 directions commonly known as roll, pitch and yaw angle. Quadcopters can fly in two configurations (see Figure 1). The first is the 'Cross-flyer' configuration. In cross flyer formation, the rotors of the copter are aligned along the roll and pitch axes whereas the second configuration is known as the 'X-flyer' configuration having a pair of frontal rotors and rear rotors. Both configurations have advantages and disadvantages. In cross-flyer configuration the roll and pitch angles, the forward pitch is controlled by only one rotor whereas in the 'X-flyer' configuration the forward pitch is controlled by a pair of rotors. "X-Flyer" performs better when performing displacements. In [10] the authors have shown this comparison using experimental data. In addition, another important reason for choosing the "X-Flyer" configuration as more adequate to the research objectives is because UAV vision systems are commonly installed pointing downwards and forwards. Considering the vision system pointing forwards for performing search operation, obstacle detection and avoidance it is desirable to have a clear frontal view of the quadcopter.

b) <u>UAVs Flying Strategy</u>: Let's discuss how a quadcopter flies in a 3-dimensional environment. The dynamic model [10-17] of the quadrotor can be obtained through the Euler-Lagrange approach or the Newton-Euler approach. The position of a quadcopter can be expressed as a vector with 6 quantities like below;

$$q \; = \; (x, \; y, \; z, \; \psi, \; \theta, \; \varphi) \in R^6$$

where $\xi \; = \; (x, \; y, \; z) \in R^3$ is the position vector of the quadcopter (centre of mass of the quadcopter) relative to a fixed inertial frame (here it is the ground station or base station). The rotorcraft's Euler angles (the orientation of the rotorcraft) are expressed by $\eta \; = \; (\psi, \; \theta, \; \varphi) \in R^3$, $\psi$ is the yaw angle around the z-axis, $\theta$ is the pitch angle around the y-axis and $\varphi$ is the roll angle around the x-axis [13].

In normal condition, two-rotors rotate in a clockwise direction and the other two rotors rotate in an anti-clockwise direction with the same speed. So, the torque generated in clockwise and anti-clockwise directions balances each other. Now, if the torque generated by these two pairs of rotors are different from each other then the quadcopter will spin about the vertical axis. The rotating wings generate air thrust which is proportional to the square of the angular velocity of the wings. If the sum of the air thrust is equal to its weight then it will hover otherwise it will move in the vertical direction. The forward or backward movement of the quadcopter can be generated by generating forward or backward pitch with a proper pitch angle. Similarly, the quadcopter can move in right or left by generating a proper roll angle. The thrust produced by a motor Mi (for i=1, ..., 4) is

$$F_i = k \cdot \omega_i^2$$

This thrust is always in the vertical direction of the rotor. The torque generated by the motor is opposed by an aerodynamic drag by the following equation,

$$I_{rot} \cdot \dot{\omega} = \tau - \tau_{drag}$$

The aerodynamic drag is defined by,

$$\tau_{drag} = \frac{1}{2} \cdot \rho v^2$$

where p is the density of air, A is the area of the rotating face, and v is the velocity concerning the air.

$$m\ddot{x} = u(sin\,\varphi\,sin\,\psi \, + \, cos\,\varphi\,cos\,\psi\,sin\,\theta)$$

$$m\ddot{y} = u(cos\,\varphi\,sin\,\theta\,sin\,\psi \, - \, cos\,\psi\,sin\,\varphi)$$

$$m\ddot{z} = u\,cos\,\theta\,cos\,\varphi \, - \, mg$$

$$\ddot{\psi} = \tilde{\tau}_\psi$$

$$\ddot{\theta} = \tilde{\tau}_\theta$$

$$\ddot{\varphi} = \tilde{\tau}_\varphi$$

where x and y are coordinates in the horizontal plane, z is the vertical position. The first derivative of these coordinates represent the velocity and the second derivatives denote the acceleration towards the respective axes. And $\tilde{\tau}_\psi$, $\tilde{\tau}_\theta$ and $\tilde{\tau}_\varphi$ are the yawing moments, pitching moment and rolling moment, respectively, which are related to the generalised torques, $\tau_\theta$ and $\tau_\varphi$. $u$ is the total thrust acting on the quadcopter which mathematically can be represented,

$$u = \sum_{i=1}^{4} F_i$$

## 5. Control of UAVs: Control and path planning [18] are two crucial and most important tasks during the flight of a UAV. In this section, we have discussed several control approaches like PID/PD control [11, 14-15], Artificial Potential Field [19-20] etc.

A.  <u>PD Contro</u>l: PD stands for proportional and derivative. Let's suppose there is a UAV in hovering position and $x_{des}(t)$ is its desired trajectory but its current trajectory is x(t). So, here the error function is $x_{des}(t)$-x(t). Now, this is an automatic control problem where the UAV trajectory is our plant. And, we have to minimise the error function. It is better if it decreases exponentially. From mathematics, we can say if we can find some positive value of $k_p$ and $k_d$ then the error function will decrease exponentially.

$$e(t) = x_{des}(t) - x(t) \text{ --------------(9)}$$

$$\frac{d}{dt^2}[e(t)] + k_p \cdot \frac{d}{dt}(e(t)) + k_d \cdot e(t) = 0 \text{ ----------------(10)}$$

Now, here $k_p$ and $k_d$ are two positive constants. For some value, if the function satisfies the above equation. Here, $k_p$ is known as proportional constant whereas, $k_d$ is known as the derivative constant. In some cases, we add an integral term to satisfy the equation more conveniently. Then equation (10) becomes,

$$\frac{d}{dt^2}[e(t)] + k_p \cdot \frac{d}{dt}(e(t)) + k_d \cdot e(t) + k_i \int e(t)'dt = 0$$

Here, $k_i$ is known as the integral constant. $k_p$ acts as a springy constant, if its value is too high then the UAV will oscillate concerning its desired position where $k_d$ acts as damping constant. A higher value of $k_d$ will increase the time taken by the UAV for settling at its desired position.

## 6. Effect of Wind gust on UAVs:

The control of UAVs is very difficult and challenging due to the external disturbance of the atmospheric environment. It also affects the safety of the UAV as well. In the environment, there are several types of wind that affect UAVs' trajectory control and operation. It may distort the UAV structure or antenna structure [21]. Here we will discuss only three common types which affect mostly a UAV. These are;

A. Average wind: The average wind is also known as normal wind. Its speed changes with the spatial and temporal variation. It flows in a uni-direction at a constant speed.

B. Turbulence: The turbulence is also called a continuous fluctuation. The turbulence has a direct relationship with many factors such as heat exchange and wind shear. In fluid dynamics applications, turbulence can be described by stochastic theory and method. The two common models of turbulence are the Dryden model (DM) [22] and the Von Karman model (VKM) [23]. Both models are depending upon the statistics and measurements. The VKM first establishes the turbulence spectral function and then gathers the correlation function and vice versa in DM. The simplest dissimilarity between these two models is only its slope having a greater frequency of its spectral function. For solving engineering issues, both of them are used. The spectral Dryden function defined as,

$$\Phi_u(\Omega) = \sigma_u^2 \cdot \frac{L_u}{\pi} \cdot \frac{1}{1+(L_u\Omega)^2}$$

$$\Phi_v(\Omega) \; = \; \sigma_v^2 \cdot \frac{L_v}{\pi} \cdot \frac{1+12(L_v\Omega)^2}{(1+4(L_v\Omega)^2)^2}$$

$$\Phi_w(\Omega) \; = \; \sigma_w^2 \cdot \frac{L_w}{\pi} \cdot \frac{1+12(L_w\Omega)^2}{(1+4(L_w\Omega)^2)^2}$$

The Von Karman spectral function is defined as;

$$\Phi_u(\Omega) \; = \; \sigma_u^2 \cdot \frac{L_u}{\pi} \cdot \frac{1}{(1+(aL_v\Omega)^2)^{5/6}}$$

$$\Phi_v(\Omega) \; = \; \sigma_v^2 \cdot \frac{L_v}{\pi} \cdot \frac{1+(8/3)(2aL_v\Omega)^2}{(1+2a(L_v\Omega)^2)^{11/6}}$$

$$\Phi_w(\Omega) \; = \; \sigma_w^2 \cdot \frac{L_w}{\pi} \cdot \frac{1+(8/3)(2aL_w\Omega)^2}{(1+2a(L_w\Omega)^2)^{11/6}}$$

Whereas,

(i) $\sigma_u$, $\sigma_v$, and $\sigma_w$ are three directions' wind speed

(ii) $L_u$, $L_v$ and $L_w$ are three directions' wavelength of turbulent flow

(iii) $\Phi_u$, $\Phi_v$, and $\Phi_w$ are spectral functions whose directions are alongside the axis of the body coordinate system of UAV. In VKM spectral function, the value of a = 1.339.

C. <u>Wind Shear</u>: The wind shear or wind gradient is the difference between the two vectors of wind at two points distributed by the space between two points. For example, in Figure 2 the vectors at each position represent the wind directions by the directions of vectors. There are two different wind field, when a UAV flies from the A  to B or B to A, it experiences a wind shear. The main causes of wind shear during the UAV flight are frontal, nocturnal, and micro-downburst. The difference between the wind shear and turbulence is frequency. The wind shear is of low frequency, and it changes every few seconds and it has a proper direction. Sudden variation in average wind speed can produce wind shear.

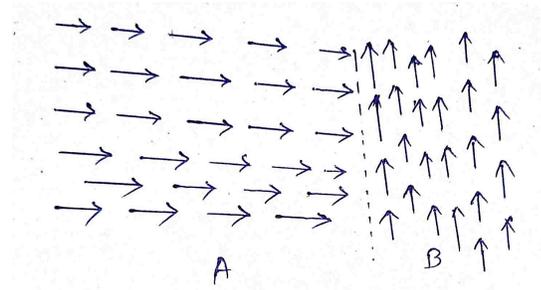

Figure 2: Wind DIrection

When a UAV is flying in the air, the change of airspeed $V_a$, ground speed $V_g$ and average wind speed $V_w$ can be obtained as given in,

$$\Delta V_g = P\Delta V_w,$$
$$\Delta V_a = (1 - P)\Delta V_w$$

Where, $V_a$ is the airspeed, the speed of UAV relative to the air. $V_g$ is the ground speed of the UAV which is the horizontal speed of the drone relative to the earth's surface. P is a coefficient. It depends upon the effective area and the mass of the UAV, and the euclidean norm or L2 norm of P ($\|P\|_2$ ) is less than 1. This coefficient P denotes how much the ground speed of the UAV is susceptible to be disturbed by the average wind. The turbulence is very random and it exerts uneven forces on all parts of the UAV. the forces on the UAV due to the turbulence are calculated from the air resistance equation:

$$F_D = \rho \cdot v^2 \cdot C_D \cdot S$$

where $F_D$ is the force of the airflow on the aircraft, $\rho$ is the air density, $v$ is the airflow speed, S is the windward area, and $C_D$ is the airflow force coefficient which depends on the Reynolds number $R_e$ of the airflow. The airflow speed can be obtained from the Dryden model (DM) and the Von Karman model (VKM). The control of UAV in wind gusts is a crucial and difficult task. There are several methods for it. In [24] the authors had proposed a novel approach called "Reject External Disturbance" where the UAV changes its forward flight to a turning state with a minimum turning radius.

# SECTION-2

## 7. Swarm Behaviour in Nature:

Swarm means honeybee. Swarming is applied to mainly the group behaviour of insects. But in a general in a broad sense, a swarm is a large number of homogenous, simple agents interacting locally among themselves and their environment with no central control to allow a global interesting behaviour to emerge. Swarm Behaviour or swarming is a collective behaviour exhibited by entities, such groups of insects [25-26], fish [27-28], birds [29], primates [30-31], humans [32] and animals. In the past decades, biologists and natural scientists have been studying the behaviours of social insects because of the amazing efficiency of these natural swarm systems. The swarm behaviour of various animals, insects are referred to with various terms such as, flocking or murmuration can refer specifically to swarm behaviour of birds, shoaling or schooling refers to swarm behaviour of fish. Similarly, herding for tetrapods such as sheep, elephant, Pride for lion, Pack for the wolf. Basically from an abstract point of view, swam behaviour is the collective motion of a large number of self-propelled entities. The main function of swarming is considered to be a reduction in the risk of predation[33]. Such social behaviour offers a better chance for survival due to communal efforts among the swarm. The swarm behaviours can be modelled mathematically and computationally. In [34] the authors have modelled the school of fishes and flocking of birds mathematically. These models develop an integrative picture of the connection between traits at the individual and group levels. In the late 80s for the first time, computer scientists proposed the scientific insights of these natural swarm systems to the field of artificial intelligence. In 1989, the expression "Swarm Intelligence" was first introduced by G. Beni and J. Wang in the global optimization framework as a set of algorithms for controlling robotic swarms [35]. In 1991, Ant Colony Optimization (ACO) [36] was introduced by M. Dorigo and colleagues as a novel nature-inspired metaheuristic for the solution of hard combinatorial optimization (CO) problems. In 1995, particle swarm optimization was introduced by J.Kennedy et al. [37].

## 8. Swarm Intelligence:

Swarm intelligence[38] models are the computational models inspired by natural swarm systems. Till now several swarm intelligence models based on different natural swarm systems have been proposed in the literature, and successfully applied in many real-life applications. Some most common swarm intelligence models are: Ant Colony Optimization [36, 38], Particle Swarm Optimization [37-38], Artificial Bee Colony [39], Bacterial Foraging [40], Cat Swarm Optimization [41], Artificial Immune System [42], and Glowworm Swarm Optimization [43], Wolf Pack Algorithm [44], Lion optimization algorithm [45-46]. Swarm Intelligence algorithms in several optimization tasks and research problems. Swarm Intelligence principles have been successfully applied in a variety of problem domains including function optimization problems, finding optimal routes, scheduling, structural optimization, and image and data analysis. Computational modelling of swarms has been further applied to a wide range of diverse domains, including machine learning, bioinformatics and medical informatics [35], dynamical systems and operations research.

Let's discuss two important swarm optimization algorithms;

A. Particle Swarm Optimization:
PSO [32, 38] is a population-based search strategy that finds optimal solutions using a set of flying particles with velocities that are dynamically adjusted according to their historical performance, as well as their neighbours in the search space. PSO solves problems whose solutions can be represented as a set of points in an n-dimensional solution space. Here, the term particles refer to the population members, which are fundamentally described as the swarm positions in the n-dimensional solution space. Each particle is set into a motion through the solution space with a velocity vector representing the particle's speed in each dimension. Each particle has a memory to store its historically best solution (i.e., its best position ever attained in the search space so far, which is also called its experience). These experiences are the key point of the success of the PSO algorithm. The particles share these experiences with their neighbour particles. At first, the PSO was designed to optimise real-value continuous problems, The original version of the PSO algorithm is essentially described by the following two simple velocity and position update equations, shown by the below two equations respectively.

$$v_{id}(t+1) = v_{id}(t) + c_1 R_1(p_{id}(t) - x_{id}(t)) + c_2(p_{gd}(t) - x_{id}(t)) \quad ---- - \ equation \ (1)$$

$$\text{And,} \quad x_{id}(t+1) = x_{id}(t)) + v_{id}(t+1) \quad ---- - \ equation(2)$$

Here, $v_{id}(t)$ is the velocity of $i^{th}$ particle in the $d^{th}$ dimension and t denotes the iteration number. $x_{id}(t)$ is the position of $i^{th}$ particle in the $d^{th}$ dimension. $p_{id}(t)$ is the historical best position reached by $i^{th}$ particle in the $d^{th}$ dimension. $p_{gd}(t)$ is the global best position of the swarm. $R_1$ and $R_2$ are two n-dimensional vectors and $c_1, c_2$ are two constants named personal learning factor and social learning factors respectively.

The PSO algorithm can be described briefly like below;
- Initialise the swarm by randomly assigning each particle to an arbitrarily initial velocity and a position in each dimension of the solution space.
- Evaluate the desired fitness function to be optimised for each particle's position.
- For each particle, update its historically best position so far, $P_i$, if its current position is better than its historically best one.
- Identify/Update the swarm's globally best particle that has the swarm's best fitness value, and set/reset its index as g and its position at $P_g$.
- Update the velocities of all the particles using equation (1).
- Move each particle to its new position using equation (2).
- Repeat steps 2–6 until convergence or a stopping criterion is met (e.g., the maximum number of allowed iterations is reached; a sufficiently good fitness value is achieved, or the algorithm has not improved its performance for several consecutive iterations)

We have discussed the original PSO algorithms. Later researchers have worked on it and improved the PSO algorithm to increase its performance and robustness. In [47] authors had surveyed almost 350+ research works on improvement PSO.

B. Wolf Pack Algorithm: Wolves have a clear social work division in their group or pack. There is a lead wolf; some elite wolves act as scouts and some ferocious wolves in a wolf pack. Now, we will discuss the coordination of wolves in a pack during praying.
Firstly, the lead wolf is responsible for commanding the wolves and constantly making decisions by evaluating the surrounding situation and perceiving information from other wolves. Secondly, the lead wolf sends some elite wolves to hunt around and look for prey in the probable scope. Those wolves are scouts. They walk around and independently make decisions according to the concentration of smell left by prey. Thirdly, once a scout wolf finds the trace of prey, it will howl and report that to the lead wolf. Then the lead wolf will evaluate this situation and decide whether to summon the ferocious wolves to round up the prey or not. If they are summoned, the ferocious wolves will move fast towards the direction of the scout wolf. Fourthly, after capturing the prey, the prey is not distributed equitably, but in order from the strong to the weak. That is to say that, the stronger the wolf is, the more food it will get.

In the WPA algorithm [48-51] there are a total of 3 artificial intelligent behaviours and two rules. These are scouting behaviour, calling behaviour, besieging behaviour and winner-take-all rule for generating lead wolf, and the strong-survive renewing rule for the wolf pack. At first, the scouting behaviour increases the possibility that WPA can fully traverse the solution space to find the optimum solution for the given problem. Secondly, the winner-take-all rule generates the lead wolf by comparing the function value of the lead wolf with the best one of other wolves in each iteration. If the value of the lead wolf is not better, it will be replaced and the best wolf becomes a lead wolf. The calling behaviour makes the wolves move towards the lead wolf whose position is the nearest to the capturing prey. As the step of wolves in calling behaviour is largest among all three behaviours, so the winner-take-all rule and calling behaviour make wolves arrive at the neighbourhood of the global optimum only after a few iterations elapsed. Thirdly, with a small step, besieging behaviour makes WPA algorithms have the ability to open up new comparably small solution space and carefully search the global optima in that small solution area. Fourthly, with the help of stronger-survive renewing rules for the wolf pack, the algorithm can get several new wolves whose positions are near the best wolf, which allows us to search the global optimum more efficiently. All the above make WPA possess superior performance in accuracy and robustness. Now, let's discuss the WPA algorithm briefly.

❖ Initialise the following parameters, the initial position of artificial wolf $i$ ($X_i$), the number of the wolves ($N$), the maximum number of iterations ($k_{max}$), the step coefficient ($S$), the distance determinant coefficient ($L_{near}$), the maximum number of repetitions in scouting behaviour ($T_{max}$), and the population renewing proportional coefficient ($\beta$).

❖ The wolf with the best function value is considered a lead wolf. In practical computation, $S_{num} = M_{num} = n - 1$, which means that wolves except for the lead wolf act with different behaviour as different statuses. So, here, except for the lead wolf, according to the scouting behaviour, the rest of the $n - 1$ wolf firstly act as the artificial scout wolves to take scouting behaviour until $Y_i > Y_{lead}$ or the maximum number of repetition $T$ max is reached and then go to Step 3.

❖ Except for the lead wolf, the rest of the $n - 1$ wolf secondly act as the artificial ferocious wolves and gather towards the lead wolf according to calling behaviour; $Y_i$ is the smell concentration of prey perceived by wolf $i$; if $Y_i \geq Y_{lead}$, go to Step 2; otherwise, the wolf $i$ continues running until $L(i, l) \leq L_{near}$; then go to Step 4.

❖ The position of artificial wolves who take besieging behaviour is updated according to besieging behaviour.

❖ Update the position of the lead wolf under the winner-take-all generating rule and update the wolf pack under the population renewing rule according to (6).

❖ If the program reaches the precision requirement or the maximum number of iterations, the position and function value of the lead wolf, the problem optimal solution, will be output; otherwise, go to Step 2.

C. <u>Grey Wolf Algorithm</u>: Grey wolves belong to the Canidae family. They follow a strict hierarchy of social dominance [52]. They live in a pack of 5-12 members on average. They have a total of four categories of wolves in their pack. These are alpha, beta, delta, and omega. The alpha is the best in terms of managing the pack. Beta is the lower level of alpha, they are bound to report alpha. Delta is at the lower level of beta and omegas are at the lowest level. All top 3 levels can command these omegas. In the GWO algorithm [52-54] the best solution is marked as the α; the second-best solutions are marked as the β, the third-best solutions are marked as the δ, the rest of the solutions are marked as ω. According to Muro et.al [55], the main phases of grey wolf hunting are as follows:

● Tracking, chasing and approaching the prey.
● Pursuing encircling, and harassing the prey until it stops moving.
● Attack towards the prey.

<u>Encircling prey</u>: The grey wolves have an intrinsic functionality of hunting around the prey.

$$D = | \ C \ . \ X_P(t) - X(t) \ | \ \text{------------------(3)}$$

$$X(t + 1) = X_P(t) - A \cdot D \ \text{-----------------(4)}$$

The distance between the prey and the wolf here is D. X Is the wolf's vector of position and $X_P$ implies beheads position vector at the iteration t. A and C are random vectors that are determined in Eqs. (3) and (4).

$$A = 2a \cdot r_1 - a \ \text{-----------------(5)}$$
$$C = 2 \cdot r_2 \ \text{-----------------(6)}$$

The random vectors in the range of [0, 1] here are, $r_1$ and $r_2$. They make wolves reach among the prey and the wolf at any point. Vector works to regulate the GWO algorithm phenomenon and is used as a basis for A computations. The vector aspect standards decline dynamically among 2 and 0 overtime.

Prey hunting: The grey wolves can feasibly circle it because they can track the movement of the prey. The ( α ) wolf tends to lead nearly the entire process of hunting. All grey wolves are chased by ( α ), ( β ), and ( δ ) wolves. They will also keep updating their positions to the optimal position of the wolves ( α ), ( β ), and ( δ ). It is articulated in Eqs. (8)–(10) in statistical terms.

$$D_\alpha = |C_1 \cdot X_\alpha - X|, \ D_\beta = |C_2 \cdot X_\beta - X|, \ D_\delta = |C_3 \cdot X_\delta - X| \text{----------(7)}$$
$$X_1 = X_\alpha - A_1 \cdot D_\alpha, \ X_2 = X_\beta - A_2 \cdot D_\beta, \ X_3 = X_\delta - A_3 \cdot D_\delta \ \text{-------- (8)}$$

Eq. (7) can be used to measure the upgraded status of the grey wolf.

$$X(t + 1) = \frac{X_1 + X_2 + X_3}{3} \text{-----------------(9)}$$

Prey searching and attacking: Grey wolves will only attack the prey when they are no longer moving.

The pseudocode of the Grey Wolf Algorithm is here;
- Initialise the grey wolf population $X_i$ (i=1,2,......,n).
- Initialise a, A, and C.
- Calculate the fitness of each search agent. $X_\alpha$ is the best search agent, $X_\beta$ is the second-best search agent and $X_\delta$ is the third-best search agent.
- While (t<MAX_ITERATION)

     For each search agent

         Update the position of the current search agent by using eq. (9)

     End for

     Update a, A, and C

     Calculate the fitness of all search agents

     Update $X_\alpha$, $X_\beta$ and $X_\delta$

     $t = t + 1$

  End while

- Return $X_\alpha$

# 9. Swarm in UAVs:

We have already analysed in the introduction part the challenges faced by a single drone. To overcome these challenges the most common and versatile solution is a group or combination of UAVs or in terms of a swarm of UAVs. A swarm of UAVs is much more helpful than some of the single UAVs. Lav Gupta et. al, have beautifully discussed in their work [6], the benefits of swarm UAV networks over single drone networks. Some of these benefits are enlisted in Table 1.

| Feature | Single UAV | Swarm UAV |
|---|---|---|
| Chances of mission failure | High | Low |
| Scalability | Limited | High |
| Survivability | Poor | High |
| Speed of Mission | Slow | Fast |
| Cost | High | Low |
| Bandwidth Required | High | Medium |
| Complexity | Low | High |

Table 2: Comparison between Single UAVs & Swarm UAVs

However, to get these benefits from swarm UAVs we have to tackle various issues like changing the topology of the swarm dynamically, mobility, power constraints, collision avoidance. In terms of communication needs, the UAVs should communicate with each other with very low latency to keep minimal distance among them and avoid a collision. As the number of UAVs increase the challenges also increase.

# 10. Swarm Architecture in UAV networks:

The swarm architecture in UAV networks plays an important role in the control, communication, and autonomous collaboration of UAV swarms. In [56-60] researchers had highlighted various types of UAV networks.

A. Centralised UAV Network: Centralised UAV topology can be classified into two major categories.

a. <u>Star Network</u>: Here each UAV can communicate with the ground station directly but they have to be routed through the ground to communicate with each other. As the ground station is the central node, the system isn't robust. This network is relatively simpler. It is suitable when the number of UAVs and coverage is small. But, the reliability of this network is very low. It suffers from the disadvantages of high latency, single point of failure (SPOF). A centralised UAV network architecture has shown in Figure 3.

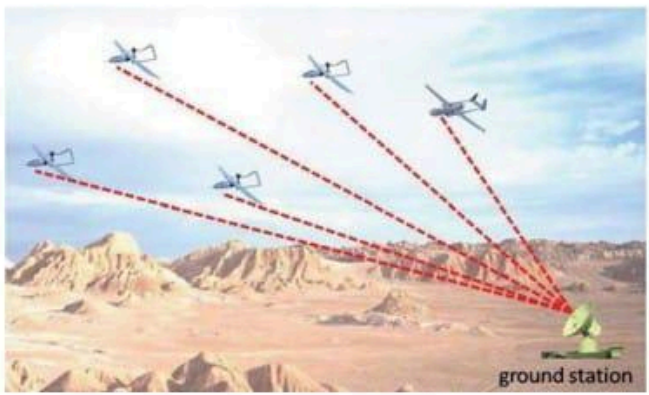

Figure 3: Star Network [60]

b. <u>Multi Star UAV Network</u>: In multi-star networks, a group of UAVs is connected with a master UAV in a decentralised manner and some master UAVs are connected with the ground control station in a decentralised manner.

B. <u>decentralised UAV Network</u>: The decentralised UAV networks can be classified into 3 major categories;

a. <u>Single Group UAV Ad Hoc Network</u>:  Here one master UAV is connected with the ground station and this master or backbone UAV serves as a gateway of the ad hoc network. Its coverage area is larger than centralised architecture and also it is more power-efficient.  The intra-swarm communication can be done in various architectures like rings, mesh, trees, or fully connected. The UAV Ad Hoc network is more robust than the centralised configuration. It also has much more stability, flexibility. But this network may suffer from scalability problems [56]. The architecture is shown in Figure 4.

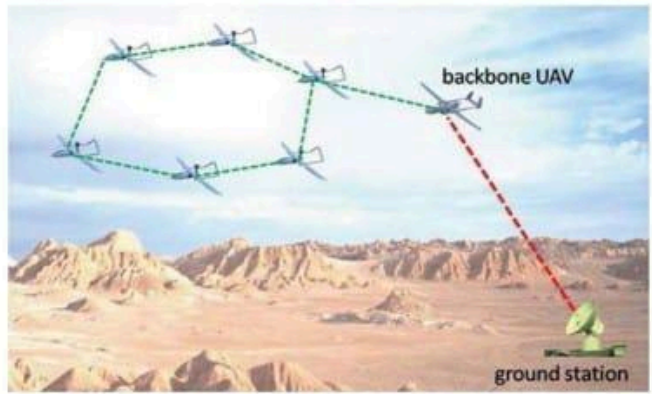

Figure 4: Single Group Ad-Hoc Network [60]

b. <u>Multi-Group Ad-Hoc Network</u>: Here UAVs within a group form a UAV ad hoc network with its respective master UAV. Intra-group communications are performed through the backbone UAV of each group whereas, communication between two groups is performed through the ground station. Due to its semi-centralized structure, it has some lack of robustness. The architecture is shown in Figure 5.

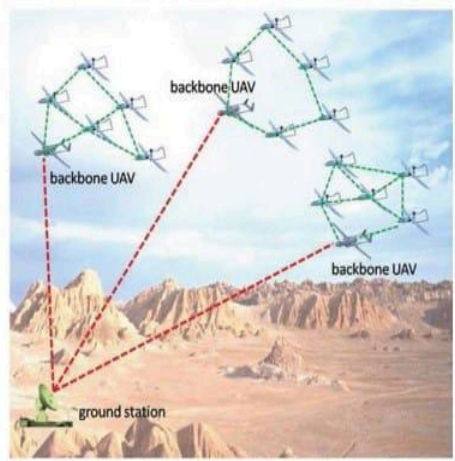

Figure 5: Multi-Group Ad-Hoc Network [60]

c. <u>Multi-layer Ad-hoc network</u>: Here UAVs within a group form a UAV ad hoc network and the backbone UAV of each group form another layer of the UAV Ad-hoc network. Information exchange between two UAVs of a group is performed through the backbone UAV of that group. In this architecture information exchange between any two UAV groups does not need to be routed through the ground station. So, this network is robust. The architecture is shown in Figure 6.

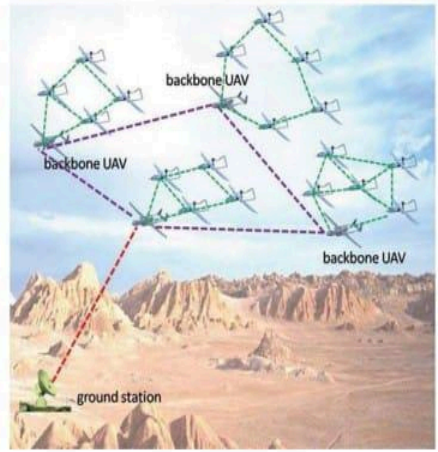

Figure 6: Multi-Layer Ad-Hoc Network [60]

# 11. Communication in UAV Swarm: In a UAV swarm mainly two
communication techniques [59] are being followed for communicating in between swarm
UAVs.

    A. <u>Routing Technique</u>: In the routing technique for communication in between
      source and destination (let's suppose slave and master) the information is
      propagated along the best optimal path by node to node (UAVs are termed as
      nodes). The best optimal is chosen by various graph algorithms like the Dijkstra
      algorithm, Bellman-Ford, Floyd-Warshall Algorithm, Johnson's Algorithm, A-star
      algorithm, etc. Here, the UAV network is considered as a graph with the UAVs as
      nodes and communication links between them as edges. In [61] the authors have
      surveyed the performance of some of these algorithms.

    B. <u>Flooding Technique</u>: Unlikely the routing technique, in the flooding technique the
      message propagates through all the available nodes in the network. It is very
      simple and highly reliable. There is no sophisticated algorithm, no need for
      network management, no need for self-discovery, self-repair algorithms. These
      features make the flooding technique much easier.

<u>Comparison between Routing and Flooding Techniques</u>: In the routing technique, the
message propagates through fewer nodes which makes the system energy efficient.
However, this technique suffers from reliability and stability problems due to a single
point of failure (SPOF). A self-healing algorithm is highly required to ensure the
continuous flow of information from the source node to the destination. Determining the
routing path every time makes the communication strategy suffer from latency. Wherever
the flooding technique is free from these problems. The flooding technique has a much
better range than the routing technique. However, the main challenge with flooding
techniques is the synchronisation of information at each node. Using a synchronised flooding technique using a synergic combination of techniques solves these challenges. For networks of small sizes, the power consumption is low in routing technique, whereas for comparatively larger networks the power consumption is low in flooding technique.

<p style="text-align:center;">SECTION-4</p>

In this section, we are going to focus on existing path planning techniques and leader-follower strategies for the swarm of UAVs.

## 12.  Existing Path Planning Techniques:

There is a lot of research work [62-75] currently going on UAVs path planning. Several techniques are available for UAV path planning. Here we will discuss a few of them which are most common. During the comparison of the UAV path planning algorithm, we look at several things like its optimality, time efficiency, cost, energy efficiency, stability, complexity and many more. The path planning algorithm mainly consists of three steps. These are;

- Environmental modelling using geometrical shapes from the knowledge of the environmental map.
- Task modelling from the modelled environment.
- Path search.

At first, the obstacles are modelled as geometrical shapes from the information given in the environmental map. Like the buildings can be considered as cubes or polyhedrons, electrical poles or towers can be considered as cylinders. Normally for offline path planning, environmental modelling is done before the path planning. But, in online planning UAVs or mobile robots do not model the whole environment, it just senses obstacles and avoids those.

The second step of path planning is task modelling from the map modelled in the first step. For task modelling, there are several well-known approaches. Some of them which are comparable older are mainly graph-based where a graph is constructed between the starting point and the destination point. Recently tree-based approaches are becoming more popular for their fast convergence.

In the third step, we have to search a path from the source to the destination. There are a lot of algorithms that exist for path search. Now, we will discuss a few of the existing and popular strategies for path planning i.e. environmental modelling, task modelling and path search.

In general, sampling-based methods are very useful for environmental modelling. There are several methods for environmental modelling under sampling-based methods. These are;

1. Cell Decomposition: In this approach, the workspace is divided into a collection of cells so that the path can be calculated easily either between the same cell or an adjacent cell. The shape of the cells can be simple rectangular or hexagonal or any other shape.
2. Voronoi Diagram: Voronoi diagram divides the surface into regions based on the distance to the waypoints as shown in the below figure. The main difference between cell decomposition and Voronoi diagram is that in cell decomposition all the cells are of the same shape but in Voronoi diagram.

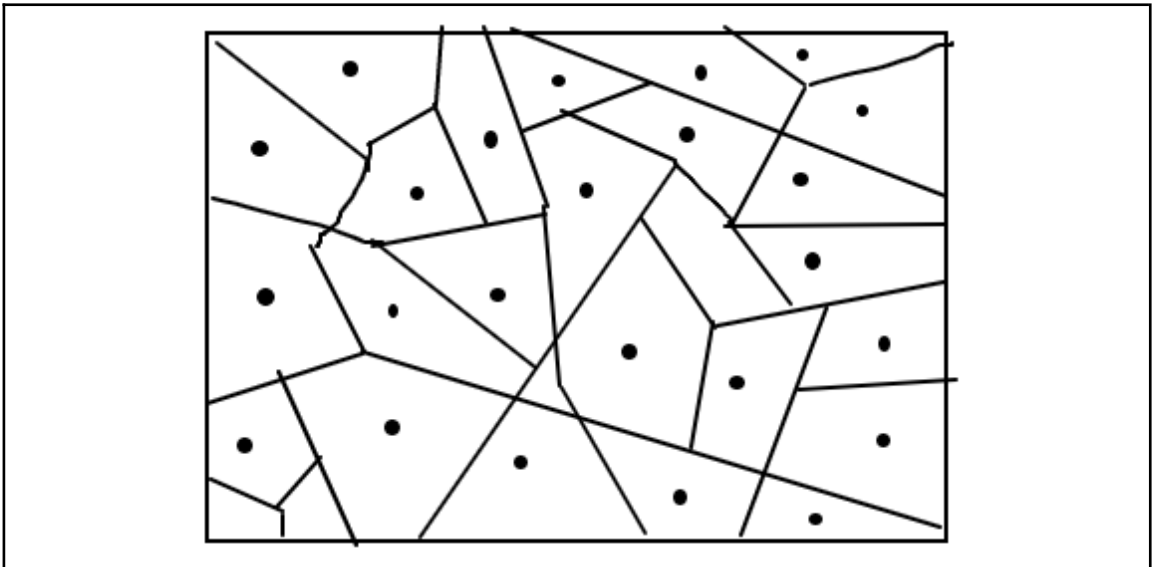

Figure 7. Voronoi diagram

3. Waypoint Graph: Waypoint graph is widely used for pathfinding [6] in the field of robotics. Each node in such a graph is called a waypoint. The waypoints are normally generated at a distance from the edges of the obstacle.
4. Roadmaps: The roadmaps method consists of two phases (i) construction and (ii) query. In the construction phase, the connectivity of the environment is computed by defining the network curves in the 3-D environment. After the construction of roadmaps, source and destination configuration points are solved in the query phase. The most commonly used algorithms for roadmaps are RRT (Rapid exploring rapid tree), PRM (Probabilistic Roadmap) etc. In some research work, researchers consider Voronoi diagram approaches, A* algorithm also roadmaps.
5. Potential Field Method: Artificial Potential Field (APF) method is a very simple but effective approach for UAV path planning. This was first proposed by Khatib [20]. In the potential field method, an attractive force from the destination and repulsive force from the source and obstacles are calculated. And then using the resultant force the trajectory is generated.

6. <u>Node-Based algorithms</u>: In a sampling-based approach, the environment is divided into several nodes. Passive roadmaps algorithms like PRM cannot pick a way out by themselves, hence a complementary search algorithm is required to find an optimal path for the UAV. Here, the environment is considered as a graph where the nodes are waypoints and edges between the nodes are the paths. The optimal and shortest path can be found using the popular graph algorithms like BFS (Breadth-First Search), DFS (Depth First Search), Dijkstra's algorithm, A* [63], D* (D* means dynamic A*), D*-Lite, Theta* etc. The node-based algorithms use dynamic programming or the greedy method.

7. <u>Bio-inspired algorithms</u>: Bio-inspired algorithms mimic the biological behaviour to deal with problems. The most common bio-inspired algorithms are evolutionary algorithms. In some research work, the researchers also consider artificial neural networks (ANN) as bio-inspired algorithms but we will discuss neural networks under AI-based approaches. Evolutionary algorithms start by selecting randomly feasible solutions mimicking the behaviour of animals, insects or any natural identities to optimise the path. A lot of Bio-inspired algorithms are available in the market. Some of these are; ACO (Ant Colony Algorithm), Genetic Algorithm (GA), PSO (Particle Swarm Optimization), MA (Memetic Algorithm), etc.

8. <u>Artificial Intelligence-based approach</u>: Recently Artificial Intelligence-based approaches [73] have geared the path planning algorithms of UAVs. Artificial Intelligence based approaches can be classified into several categories:

   a. <u>Supervised Learning</u>: When there is a relationship between the input and output data, supervised learning is the most common technique used for mapping the input and output. Artificial Neural Network (ANN) is the most popular type of supervised learning. ANN tries to adapt the behaviour of human neurons. SVM (Support Vector Machine), k-nearest neighbours algorithm, XGBoost are some other popular supervised techniques.

   b. <u>Unsupervised Learning</u>: Unsupervised learning tries to learn the pattern from the input data. This doesn't require any labelled data. Clustering is a popular example of unsupervised learning. It tries to learn the common pattern from the data.

   c. <u>Reinforcement Learning</u>: In reinforcement learning the agent learns by its action. In traditional reinforcement learning, there is a reward function that generates feedback after the agent takes an action. Based on the feedback, the agent updates the probability of each action and decides its next action. This is the most prominent area for robot path planning [74-75].

9. <u>Fuzzy-Logic Based Approach</u>: Fuzzy logic deals with situations that are neither completely true nor completely false. When pattern recognition problems arise in robotic tasks with more robustness and a perfect solution cannot be predicted and used to solve, it represents a partial solution. In this algorithm, the whole logic is divided into simpler blocks composed of a set of fuzzy logic rule statements intended to achieve the desired objective. The decision of motion is prepared only based on input parameters, not on the real-time situation.

# 13.   Comparison of existing path planning techniques:

In the above section, we have discussed several existing environmental modelling and path search algorithms. Here we will look at the pros and cons of these algorithms. Let's first briefly discuss the pros and cons of various environmental modelling approaches.

1. <u>RRT</u>:
   a. <u>Advantages</u>:
      i) Time complexity is low and hence it has the fast searching ability.
   b. <u>Disadvantages</u>:
      i) This works only for static obstacles.
      ii) The modelling of the environment is not optimal.

2. <u>PRM</u>:
   a. <u>Advantages</u>:
      i) Appropriate for complex environments
   b. <u>Disadvantages</u>:
      i) This works for static obstacles.
      ii) The solution is not optimal.
      iii) PRM is based on the concept of random trees and hence the checking for collision avoidance is expensive in terms of computation cost.

3. <u>Voronoi Diagram</u>:
   a. <u>Advantages</u>:
      i) The generated path is highly safe and has no chance of collision with the static obstacles.
      ii) Easy implementation is an advantage of this technique.
   b. <u>Disadvantages</u>:
      i) The modelling of the environment using the Voronoi diagram is incomplete.
      ii) Voronoi diagram has a converging problem.
      iii) It works only for static obstacles.
      v) This is a 2D path planning and can't work in a 3D situation.

4. <u>Artificial Potential Field (APF)</u>:
   a. <u>Advantages</u>:
      i) This is very simple and a local path search algorithm.
      ii) Implementation of APF is easier than others.
      iii) Highly effective for collision avoidance.
      iv) Its convergence rate is fast.
   b. <u>Limitations</u>:

i) APF is more suitable for static obstacles rather than dynamic obstacles.
ii) APF easily gets trapped in local minima. Practically we can avoid these local minima by considering an imaginary obstacle and then considering a repulsive force that imaginary obstacle.
iii) It is a local search technique. So, we can't rely only on APF for larger travelling distances.
iv) Theoretically it works only for static obstacles but practically it works fine for dynamic obstacles also up to some extent.

5. Dynamic Programming:
   a. Advantages:
      i) Dynamic programming considers all possible solutions and then finds an optimal solution.
      ii) It breaks the large problem into subsets which makes the computation easier.
   b. Limitations:
      i) The time complexity of dynamic programming is $O(n^2)$. Hence, when the number of nodes increases the time complexity increases.
      ii) The space complexity of dynamic programming is $O(n)$. Hence, it requires memory to save all the solutions.

6. Dijkstra's Algorithm:
   a. Advantages:
      i) Dijkstra's algorithm finds the shortest distance between the source node and destination node.
      ii) It uses a greedy strategy for path planning instead of dynamic programming. Hence, the limitations of dynamic programming can be avoided up to some extent.
      iii) The time complexity of Dijkstra's algorithm is O(ElogV), where V is the number of nodes and E is the number of edges. Hence, here computational time is less than dynamic programming.
   b. Limitations:
      i) Computational time is still high for a large number of nodes.
      ii) Space complexity is O(V+E) where V is the number of nodes and E is the number of edges that is still high.

7. Floyd-Warshall algorithm:
   a. Advantages:
      i) Floyd-Warshall algorithm finds the shortest path between all the available pairs of nodes instead of just source and destination nodes.
      ii) It performs better than Bellman-Ford and Dijkstra's algorithm in practical scenarios [73].
   b. Limitations:

i) Time complexity is $O(V^3)$ which is high when the number of nodes is very high.

ii) It finds the shortest path between all algorithms. Hence it performs a lot of unnecessary computations which are not required.

8. <u>A* algorithm</u>:
   a. <u>Advantages</u>:
      i) It can search in a huge area.
      ii) It finds the shortest between two given pairs of nodes.
      iii) It saves a significant amount of computation time
      iv) Fast searchability.
      v) On-line path searching is possible.
   b. <u>Limitations</u>:
      i) High time burden.
      ii) It works only for static obstacles.
      iii) The generated trajectory is not smooth.

9. <u>D* search algorithm</u>:
   a. <u>Advantages</u>:
      i) D* algorithm is the improved version of A* search algorithm.
      ii) It works in a dynamic environment also.
   b. <u>Limitations</u>:
      i) It uses unrealistic distance in its graph which reduces its efficiency.

10. <u>Bio-inspired Algorithms</u>:
    a. <u>Advantages</u>:
       i) These are heuristic approaches.
       ii) These algorithms can deal with a lot of variables. Hence, they can optimise multiple things at the same time.
    b. <u>Limitations</u>:
       i) Algorithm complexity and computation time is very high.
       ii) It faces premature convergence.
       iii) The applications of evolutionary algorithms are more like the trial-error method. We can't determine which evolutionary algorithm will work better at which condition.
       iv) This is an offline path searching algorithm.

11. <u>Artificial Neural Network</u>:
    a. <u>Advantages</u>:
       i) This is stable under sudden changes.
       ii) If we use transfer learning using pre-trained models then computational time will be less.
    b. <u>Limitations</u>:

i) Training time is very high. Also, depending on the size of training data and network complexity, the computational cost sometimes may be too high.
ii) Artificial neural networks require a lot of training data.
iii) The ANN is highly data-dependent. If we train the network with the data of one type of environment, then it will not perform properly in a different type of environment.
iv) As we know both input and output in supervised learning, hence the learning is not real-time.

12. <u>Unsupervised Learning</u>:
    a. <u>Advantages</u>:
        i) Unsupervised learning doesn't require labelled data.
        ii) The learning is real-time.
    b. <u>Limitations</u>:
        i) Training process is very slow.
        ii) The accuracy is quite low.
        iii) It is very uncertain to say what the algorithm is going to learn from the data.

13. <u>Reinforcement Learning</u>:
    a. <u>Advantages</u>:
        i) The best thing in reinforcement learning is that it can correct its own mistakes based on the feedback received from the environment.
        ii) It learns from its own experience.
    b. <u>Limitations</u>:
        i) The design of the reward function is not easy.
        ii) As it learns from scratch by its action, hence a lot of data and computation power is required for proper learning.
        iii) As it learns from scratch, hence the training time is too slow.

14. <u>Fuzzy-Logic Based Approach</u>:
    a. <u>Advantage</u>:
        i) Fuzzy logic is that it produces better results than a human can produce in a short period of time.
        ii) It is well suited for implementing a solution in the complex autonomous mobile system
    b. <u>Disadvantage</u>:
        i)  This doesn't provide a proper solution, always gives a partial solution. Hence, we can't rely upon only this technique.
        ii)  This is an offline algorithm and can't provide a real-time solution.

# 14.  Leader-Follower Strategy:

Drone formation studies have different approaches and objectives. The methodology chosen here for drone formation control is known as Leader-Follower [76]. It consists of an autonomous follower robot that plans its actions from the actions of a leader robot. In a multi drone system each drone has a neighbour or reference point for its movement, where the neighbour is called the leader and that particular drone which follows the reference point is called the follower. There may be several leaders and followers pairs which adds immense flexibility in case of complex arrangements or large numbers of drones.

Obviously, the leader and follower will adhere certain control measures in order to coordinate, the control has been broadly classified into two categories:
- Horizontal control: Controls the latitude and longitude.
- Vertical control: Controls the altitude

In this document we have discussed only horizontal control..
For control, two parameters are required first the relative distance between the two and second the relative angle between the two for that we need to know at least the linear and angular velocity of the leader drone.

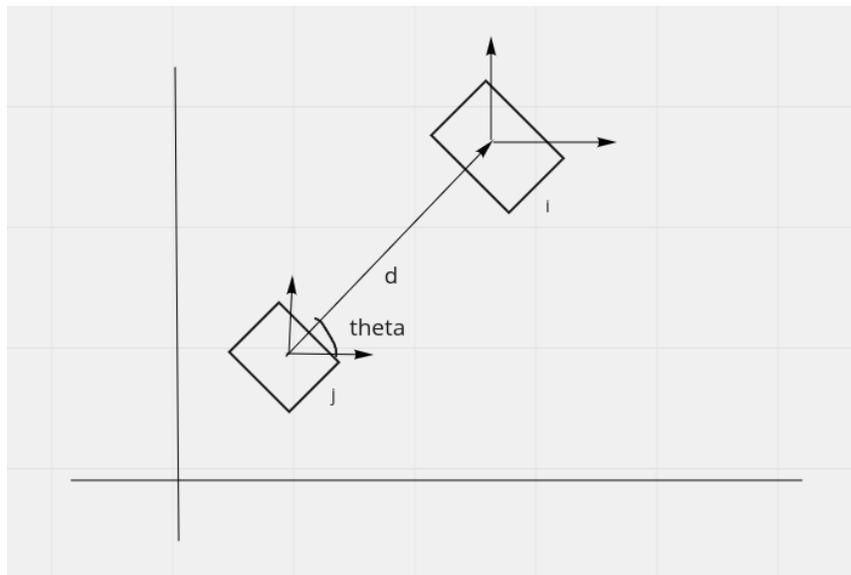

Figure 8. Horizontal schema of Leader-Follower approach

The control architecture is a multi layered architecture consisting of 4 significant layers.
1. <u>Control Layer</u>: It decides the linear and angular velocities/accelerations (v,a,ω,∝ ) of each drones and thus this layer is embedded in each and every drone.
2. <u>Movement Layer</u>: Decides the goal position and orientation of a drone. It is ordered by the Formation Layer .After analysing the positions of followers and leaders it orders the control layer.There are two mechanisms to implement this layer. Normally embedded in the following drones.

3. <u>Formation Layer</u>: It will receive commands from the Application layer and will define which drone will follow which one and how it should follow it. Also helps in transition from one formation to another. It is preferably embedded outside the formation or under formation leader.

4. <u>Application Layer</u>: Defines what kind of formation will take place and also defines the movement of the formation leader. Normally embedded outside the formation or under formation leader.

As the movement layer contains the algorithm and parameters needed for goal position and orientation, the leader-follower technique is divided into two methods based on movement. Before this, one needs to know that there exists response time (time between order received by the Formation Leader and each drone start moving) and time of formation that plays a pivotal role in variation of time taken for each and every drone in coordination.

1. <u>Fixed Global Difference</u>: The method in which the follower maintains a fixed global position/distance from the leader. Considering an X, Y and Z coordinate system, given the position of leader (XL; YL;ZL), the follower should reach the target position (XL + Δx; YL + Δy;ZL + Δz) , where Δx, Δy, and Δz are constants previously defined to indicate follower position relative to the leader.

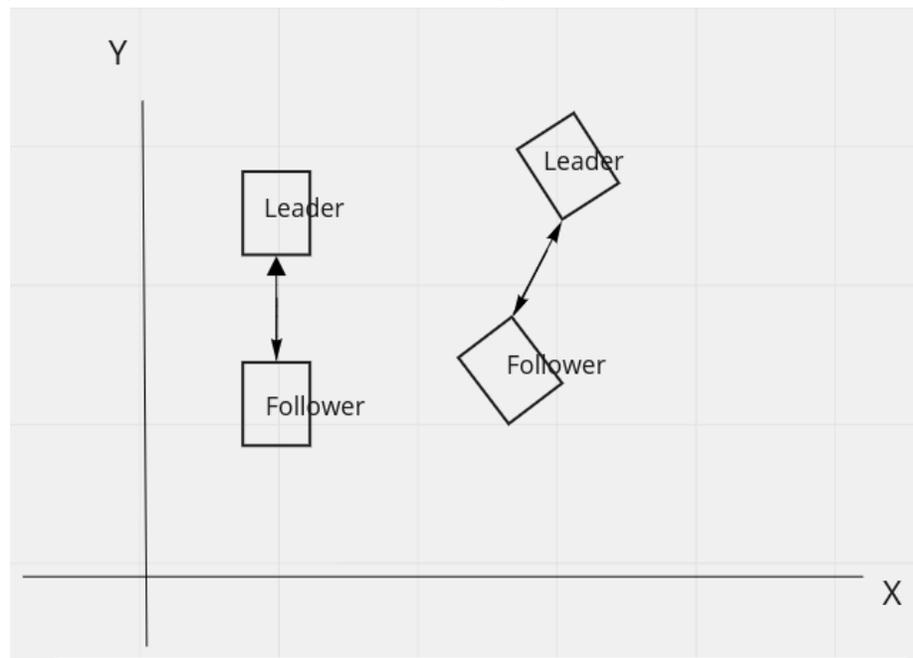

Figure 9. Fixed Global Difference (FGD) Example

2. <u>Double Fixation</u>: Approach in which the follower acts to have fixed the leader position relative to the follower, that is, the follower seeks to "see" and "be seen" by the leader in a predefined position. It defines a position (XD; YD;ZD) and orientation (θD) that the follower must maintain in relation to the leader. Given leader position and orientation PL = (XL; YL;ZL; θL) and the desired relative position and orientation (XD; YD;ZD; θD), the followers should reach the target position PO = (XO; YO;ZO; θO).

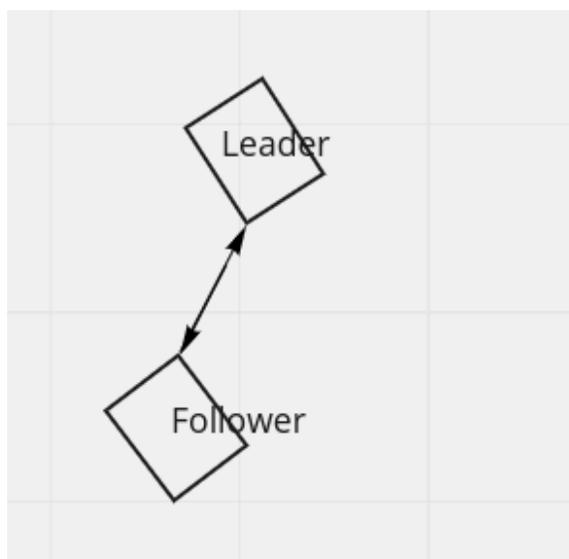

Figure 10. Double Fixation (DF) Algorithm Example

<span style="color:green">SECTION-5</span>

Under this section we are going to discuss various kinds of channel modelling and the performance of those channels i.e, path loss, bit error rate etc.

15. **Path Loss in Wireless Communication:** In a Multi-drone swarm, the slave drones communicate with the master drone in LOS (Line Of Sight) communication. In LOS communication there are mainly two models for calculating the path loss [77] of wireless communication systems. These are;

A. **Friis Model:** The Friis free space propagation model [78-79] is used to model the LOS path loss in a free space environment, considering that there is no absorption, diffraction, reflections or any other characteristic altering phenomenon like phase altering. This model states that the received power at a particular distance from the transmitter decays by a factor of the square of the distance. The Friis equation for received power is given by,

$$Pr(dB) = \frac{PtGtGr \cdot \lambda^2}{(4\pi d)^2}$$

On a log scale,

$$Pr(dB) = Pt(dB) + Gt(dB) + Gr(dB) + 20log10(\lambda) - 20log10(4\pi d)$$

B. **<u>Two Ray Ground Reflection Model:</u>** UAVs generally fly at a height of 100m during a search operation and communicate with each other through the LOS method. When an isotropic antenna is used for message transmission, it radiates into all directions including the ground. Some part of this signal gets reflected from the ground and reaches another UAV and may create a constructive or destructive pattern.[80-81]

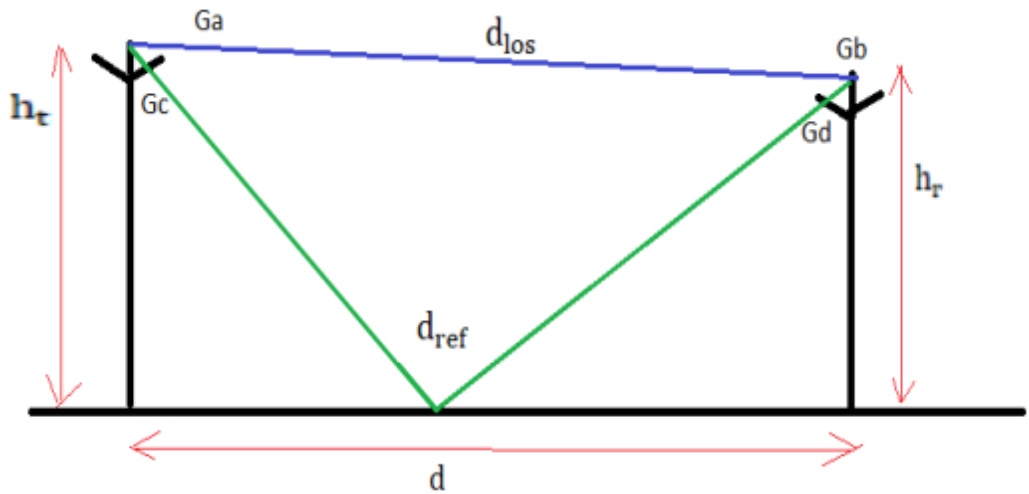

Figure 11: Two Ray Ground Reflection Model

The distance of this LOS and Reflective path is,

$$d_{los} = \sqrt{d^2 + (h_t - h_r)^2}$$

$$d_{ref} = \sqrt{d^2 + (h_t + h_r)^2}$$

At the receiving antenna, the signal will suffer a phase difference from the signal of both paths. And this phase difference will be equal to,

$$\phi = \frac{2\pi(d_{ref} - d_{los})}{\lambda}$$

Hence, the received power at the receiver antenna of the second UAV is,

$$P_r = P_t [\frac{\lambda}{4\pi}]^2 R \left[ | \frac{\sqrt{G_c \cdot G_d}}{d_{los}} + R \cdot \frac{\sqrt{G_c \cdot G_d}}{d_{ref}} | \right]^2$$

The simulation result for these two types of model is shown in the next page.

It is clear from the simulation result that due to the destructive pattern there is a huge loss in the Two Ray ground reflection model. This loss depends on both the height of the antenna of both UAVs and the distance between them. To avoid this loss the radiation pattern of the transmitter antenna should not be isotropic instead it should be directional and better if all UAVs flies in the same plane i.e., the network topology is a planner.

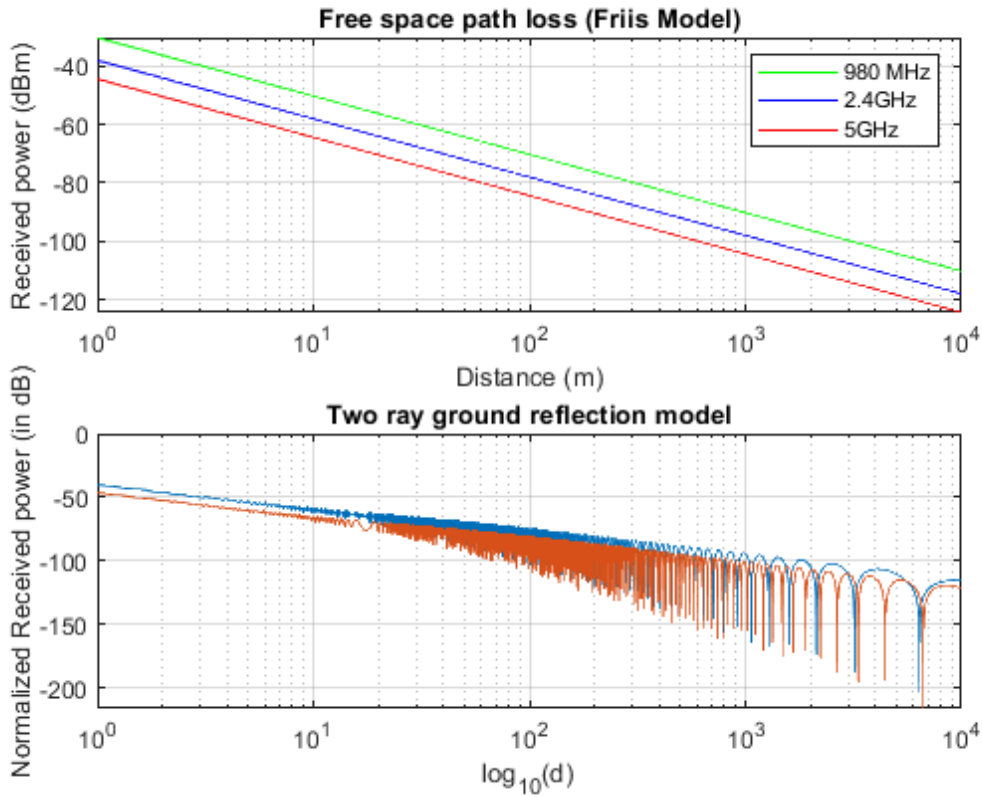

Figure 12: Received power of both models

# 16. Communication Under Noisy Channel: Here we will

discuss mainly 3 types of noisy channels. These are AWGN (Additive White Gaussian

Noise) [82], Rayleigh Channel [84] and Rician Channel [85]. Basically, Rayleigh and Rician channels cause multipath fading [83].

A. **AWGN Channel:** AWGN is a noise that affects the transmitted signal when it passes through the channel. It contains a uniform continuous frequency spectrum over a particular frequency band i.e., it has a flat power spectral density (that's why it is known as white). This noise is additive in nature and follows Gaussian distribution with zero means. The Theoretical value of Bit Error Rate (BER) of QPSK modulation [86] is,

$$BER = \frac{1}{2} \cdot erfc\left(\sqrt{\frac{E_b}{N_o}}\right)$$

The constellation diagram of the QPSK modulation of a bitstream containing 1024 bits is shown in FIgure 13.

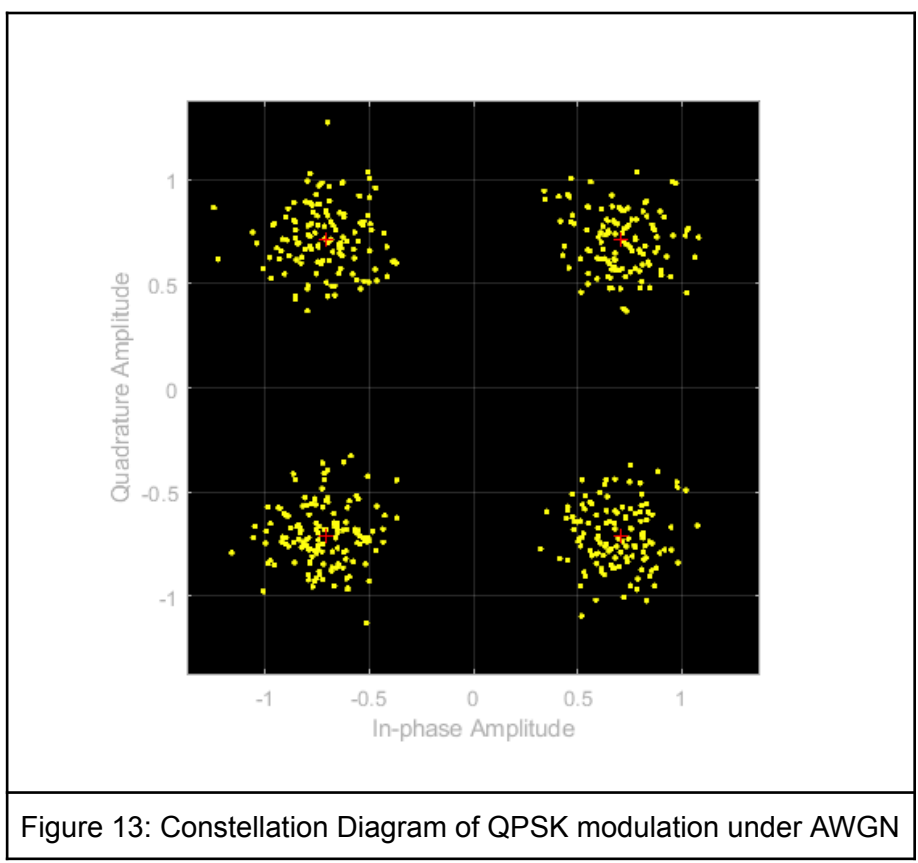

Figure 13: Constellation Diagram of QPSK modulation under AWGN

The Theoretical and simulated plot of BER with the variation $E_b/N_O$ is given in Figure 14.

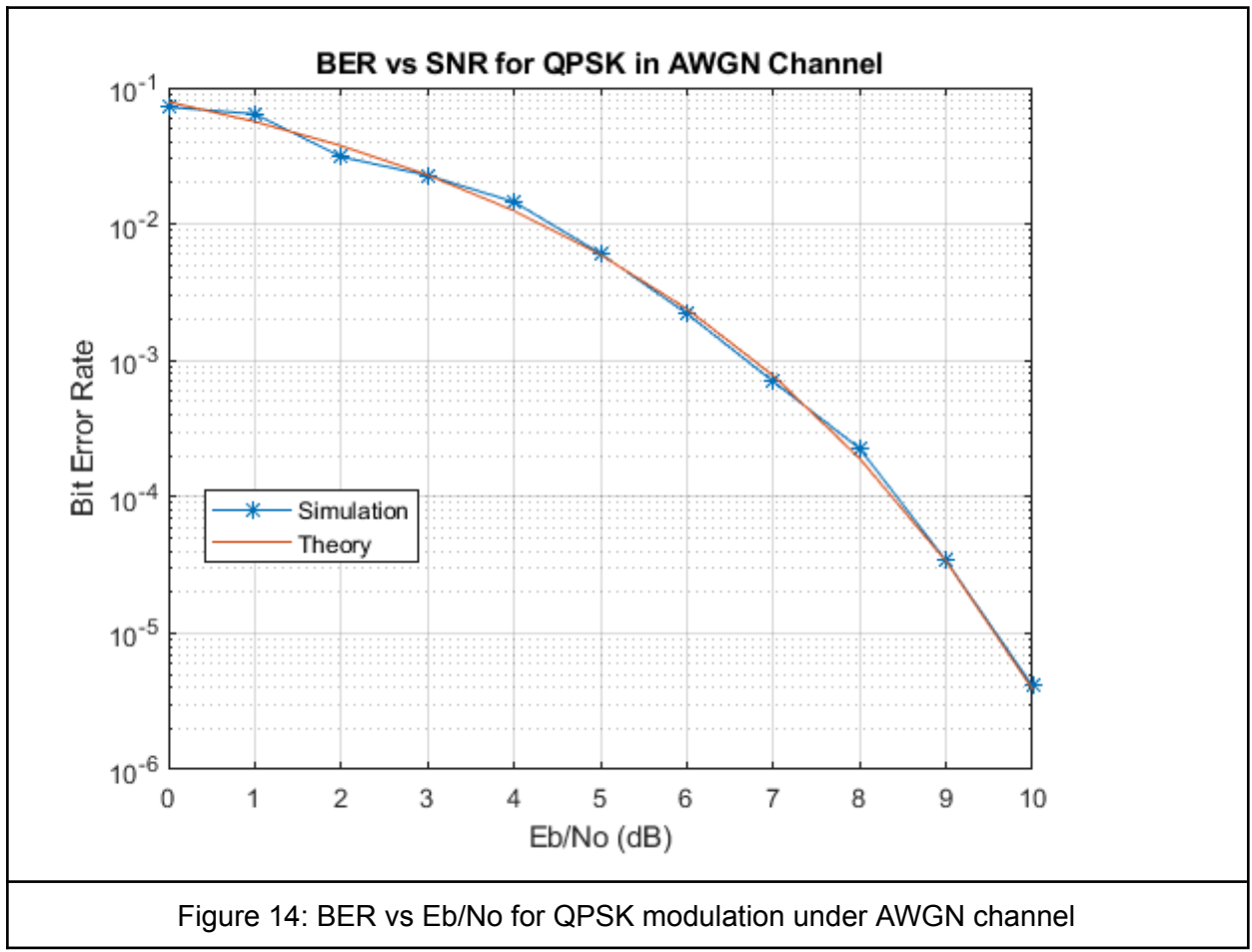

Figure 14: BER vs Eb/No for QPSK modulation under AWGN channel

B. **Rician Fading Channel**: Rician fading occurs when there is a LOS as well as the non-LOS path in between the transmitter and receiver, i.e. the received signal comprises both the direct and reflected and scattered multipath waves. Though the Rician Fading channel has a strong LOS signal, still due to the scattered multipath waves, there will be a disturbance in its constellation diagram. The constellation diagram of the Rician Fading Channel is shown in Figure 15. The BER in the Rician fading channel is quite higher than the AWGN channel. In our multi-UAV system, the communication link in between the UAVs i.e., the communication link in between the master and slave UAVs and slave UAVs itself can be modelled as the Rician Fading channel.

C. **Rayleigh Fading Channel**: Rayleigh Fading occurs when no LOS path exists in between transmitter and receiver, but only has an indirect path than the resultant signal received at the receiver will be the sum of all the reflected and scattered waves. Due to the reflected and scattered waves, there will be high BER. The constellation diagram of the QPSK modulation under the Rayleigh Fading Channel is shown in Figure 16. Rayleigh fading channel is considered as a special case of Rician fading channel where no LOS is available. The BER in this channel is much higher than the previous two. In our multi,-drone system the communication link in between the master drone and the base station may suffer the Rayleigh Fading problem and this channel can be modelled as Rayleigh fading channel.

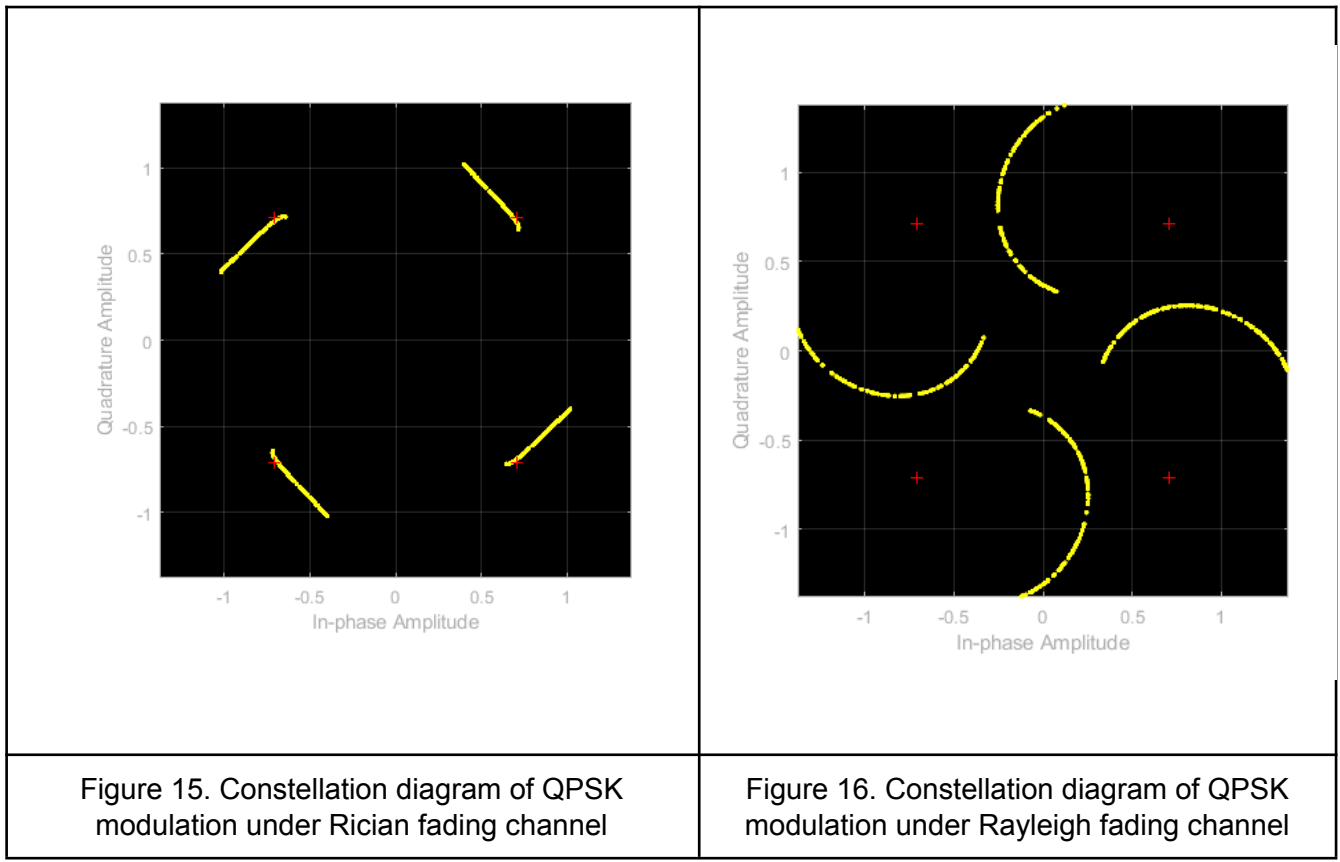

| Figure 15. Constellation diagram of QPSK modulation under Rician fading channel | Figure 16. Constellation diagram of QPSK modulation under Rayleigh fading channel |

## 17.  Link Budget Analysis: The data link communication is the crucial part of the single link communication protocol. The drone hovers high and collects the data and sends those to ground stations or base stations. The transmitting and receiving antennas are used in these aspects. Also, along with that, the drone delivers real-time flight data to the base station. Also, the base station controls the drone.

**Data Link Communication System:**

Here, we are using 2.4GHz frequency, so it's definitely in the S band. Now, to communicate with the UAV from the base station, the command-control and telemetry are important factors to be considered. The command control mechanism is held by the uplink and telemetry by downlink. In the uplink communication, the base station transmits the signal and commands the UAV and that signal is caught by the receiver part of the transceiver mounted on the UAV. The drone provides the telemetry data and the picture or video to the base station through the downlink. So, a bidirectional antenna can be used in this aspect. Due to location tracking, the GPS can be used on the UAV. That information will be sent to the base station through the downlink which is the reverse direction of the uplink.

## Payload:

The plugins on the UAV that can serve different purposes and perform different tasks are called the payload. The payload can be a camera in this case to provide the audio and video information to the base station. The UAV telemetry and the payload data is transferred to the ground station.

## Types of Antennas:

A. Dipole Antenna: In this antenna, the λ/4 electrical length is maintained. To make it mobile, the loading coil is attached at the base of the antenna. This is mainly used in the HF range. It provides very little gain.

B. Yagi Beam Antenna: It is used in directional antennas and it is a reflective type antenna.

C. Horn Antenna: It provides high gain and narrow beams. This is a type of aperture antenna. The aperture size controls the radiation pattern.

D. Reflector Antenna: It uses a parabolic reflector and has the receiver or transmitter mounted on the focal plane. It uses an L band signal and has a Low noise Block.

E. Phased Array Antennas: It provides a large ground base tracking radar and used to focus on each target in the fields, It can be used for GPS receivers.

## Antena Model Specifications:

**Model Used:** 08-ANT-0956 (Both male and female)

**Frequency Range:** 2.2 - 2.4 GHz

**Gain:** 3dBi

**VSWR:** < 1.5: 1 (For mathematical purposes, 3/2 ratio can be taken)

**Temperature**: -40 to 85

**Length of antenna :** 1.0inches

**Height:** 1.7inches

**Power:** 50 W (i/p) = $10log(50)\ dBm\ =\ 16.989\ dBm$

**Impedance (i/p)** = 50Ω

**Link length (d) :** ~2km

**Path Loss =** $-20log(\lambda/4\pi d)\ =\ -106.06dB$

**Receiver Threshold:** -85dBm

Gain (dB) = $10^{(G(dBi)/10)}\ =\ 10^{3/10}\ =\ 2dB$

**Wavelength :**

$\lambda(2.2GHz)\ =\ \frac{3*10^8}{2.2*10^9}\ =\ 0.136m$

$\lambda(2.6GHz)\ =\ \frac{3*10^8}{2.6*10^9}\ =\ 0.115m$

**Noise Figure =** $1\ +\ \frac{T_e}{T_o}$

where, $T_e$ = operational temp, $T_o$ = Standard (298K)

Therefore, F = 1 + 358/298 = 2.201

$F_{dB}$ = 20log(F) = 6.84dB

$\therefore$ Total Noise Power = $-174dB/Hz\ +\ 10log(B)\ +\ FdB$

**Reflection Coefficient, Output Impedance, Incident Power**

**VSWR** = $\frac{1+|\rho|}{1-|\rho|} \Rightarrow 1.5\ =\ \frac{1+|\rho|}{1-|\rho|}\ \Rightarrow \rho\ =\ 0.2$

$\therefore$ **Reflection Coefficient** = 0.2

$Pt\ =\ (1\ -\ \rho^2)Pi \Rightarrow 50\ =\ (1-0.04)Pi \Rightarrow Pi\ =\ 52.08W$

**Incident Power = 52.08W**

$$\rho = \frac{Z_i - Z_o}{Z_i + Z_o} \Rightarrow 0.2 = \frac{50 - Z_o}{50 + Z_o} \Rightarrow Zo = 33.33\ \Omega$$

**Output Impedance = 33.33Ω**

## LINK BUDGET

| $T_x$ | Value |
|---|---|
| $T_x$ Gain (dB) | 2dB |
| $T_x$ Loss | -0.1dB |
| $T_x$ Power | 16.989 dB |
| Radome Loss | -0.1dB |
| **EIRP** | **18.789dB** |

Table 3: Equivalent isotropic radiated power calculation

| Losses | Value |
|---|---|
| Path Loss | -101.06dB |
| $T_x$ Pointing Error | -0.5dB |
| Rain Loss | -1dB |
| Multipath | -1dB |
| Atmospheric Loss | -0.1dB |
| **Total Path Loss** | **-101.66dB** |

Table 4: Path loss calculation

| $R_x$ | Values |
|---|---|
| $R_x$ Gain (dB) | 2dB |
| Polarisation Loss | -0.1dB |
| $R_x$ Loss | -0.1dB |
| $R_x$ Pointing Loss | -0.5dB |

| Total R$_x$ Gain | 1.1dB |
|---|---|

Table 5: Receiver gain calculation

| RSL (Receiver Signal Level) | Values |
|---|---|
| EIRP | 18.789dB |
| Total R$_x$ Gain | 1.1dB |
| Total Path Loss | -101.66dB |
| **Total RSL** | **-81.171dB** |

Table 6: Received signal level calculation

| RSL | -81.771dB |
|---|---|
| Interference Margin | -1dB |
| R$_x$ Noise Figure | 6.84dB |
| Noise Bandwidth | 25MHz ( Typical value taken: Can't determine until and unless a complete frequency response is obtained) |
| **Total Noise Power** | **-93.18dBm** |
| Threshold R$_x$ | -88dB |

Table 7: Noise power calculation

**Link Margin** = EIRP + L$_{path-loss}$ + G$_{Rx}$ - TH$_{Rx}$

$\qquad$ = 18.789 + (-101.66) + 1.1dB - (-88dB)

$\qquad$ = 6.229 dB

# 18. Communication Link Performance with Distance:

The performance of communication link degrades with the increment of distance. Previously, we have observed that in wireless communication the signal power reduces with the increment of distance. Here, we are going to discuss the variation of Bit Error

Rate (BER) with link distance. There is no such straightforward relation between BER and link distance. But for each type of modulation technique, there is a relationship between BER and energy per bit. For the BPSK modulation technique;

$$BER = \frac{1}{2}\sqrt{erfc(\frac{E_b}{N_O})}$$

Here, $E_b$ is the energy per bit and $N_O$ is the noise power. Energy per bit can be calculated by dividing the signal power by data rate. And, the noise power is considered as 90 dBm. From the energy per bit and noise power we have calculated the SNR.

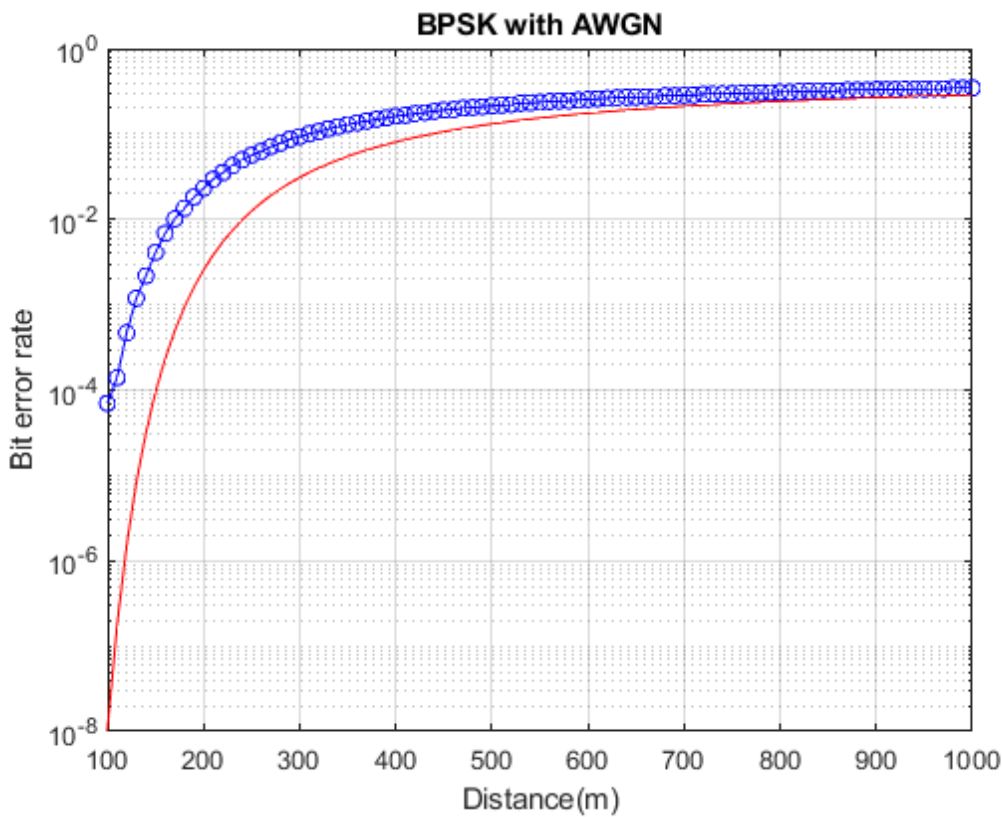

Figure 17: Variation of Bit Error Rate (BER) with distance

# 19. Improvement of Swarm UAV Communication: Swarm UAVs (Unmanned Aerial Vehicles) rely on effective communication to coordinate their actions and accomplish their tasks. Here are some ways to improve Swarm UAV communication:

- **Use of Robust Communication Protocol:** A robust communication protocol can ensure reliable communication between the UAVs and the ground control station. We have to decide and choose such a protocol which improves scalability, reduces interference, and enhances security.

- **Use of Multiple Communication Channels:** By using multiple communication channels, the UAV swarm can ensure that communication remains reliable, even if one communication channel fails. This can help prevent communication failures and improve the overall performance of the swarm. Use of multiple communication channels reduces latency and improves overall bandwidth and flexibility.
- **Machine to Machine Communication:** Instead of communicating with the ground station directly, UAVs can communicate with each other and the central UAV. So that the overall latency reduces and performance improves. Machine to machine communication will generate latency at the ground station. Hence depending on what we are trying to communicate we have to select proper communication strategy/
- **Optimization and Control of Antenna Placement:** Optimization and control of antenna placement can improve the swarm communication significantly. Antenna should be placed in such a position, so that there exists a line of sight communication among the UAVs itself. Using different types of antennas, antenna array, beamforming and changing the direction of antennas intelligently we can significantly improve overall communication performance.
- **Directional Antenna:** Instead of general omnidirectional antenna, directive antenna might be helpful to improve the communication between swarm UAVs.
- **Communication Hierarchy:** Hierarchical communication can be implemented to improve overall performances. Dedicated channels should be there for each hierarchy. The ground control station can communicate with the central UAV and then central UAV can communicate with other UAV for their actions. But, for health information like motor speed, battery life, UAVs can directly communicate with the ground control station.
- **Hybrid Communication:** Instead of using a single communication method we can use hybrid communication strategies to improve overall performance. For example, we can use LTE for ground to UAV communication whereas we can use WiFi for UAV to UAV communication.
- **Adaptive Communication Strategies:** Adaptive communication strategies are required to improve overall communication performance. Dynamic routing, multi-hop communication, adaptive power control, and cognitive radio can be used for this purpose. For example, the transmission power of the UAVs can be adjusted based on the distance between the UAVs. This can help reduce interference and improve the efficiency of the communication system. Similarly, if there is interference from other sources, the UAVs can switch to a different communication channel or re-route the communication to avoid interference.
- **Machine learning based optimization:** Machine learning algorithms can be used to predict the communication requirements of the UAVs and optimise the communication system accordingly. For example, machine learning algorithms can predict the signal strength of the communication system based on the position of the UAVs and adjust the transmission power accordingly. This can help reduce the amount of communication required and improve the efficiency of the communication system. Machine learning algorithms can analyse and predict signal strengths, analyse and predict channel performance and reroute the routing topology.

## 20. Future Work: The coverage area of the swarm of UAVs is quite large for

disaster management. And hence existing path planning algorithms are not suitable in this case due to their huge computation cost and time complexity (O(n^2) in general). Now, we are planning to focus on some vision based path planning algorithms. Traditional leader-follower strategy for maintaining the swarm structure is good during the take off or landing period. But for searching purposes it is not suitable. Because in this approach, the follower drone follows the leader and hence they will execute their search operation over the same region. So, maintaining the swarm structure is a big issue. In large networks optimization is another issue which must be explored in great detail. Efficient communication should be there in a UAV network. There is a high chance of link failure of the network in adverse situations. So, self-healing networks are highly desirable here.

## 21. Conclusion: In this work we mainly surveyed and explored a lot of topics

related to UAVs. We have found some highly promising areas for exploring in this domain. With the help of MATLAB we have shown some simulation results, especially the bit error rates during the communication between the drones. More advanced things will come in future work. Application of MIMO communication is a highly promising area for reducing the multipath fading in drone to drone communication.

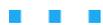